\documentclass[lettersize,journal]{IEEEtran}
\usepackage{floatrow}
\usepackage{amsmath, amsfonts, amsthm}

\usepackage{enumitem}
\usepackage{algpseudocode}
\usepackage{algorithm}
\usepackage{array}
\usepackage{textcomp}
\usepackage{url}
\usepackage{verbatim}
\usepackage{graphicx}
\usepackage{cite}
\usepackage{amsmath}
\usepackage{pifont}
\newcommand{\cmark}{\ding{51}}%
\newcommand{\xmark}{\ding{55}}%

\usepackage{pgfplots}
\usepackage{pgfplotstable}
\pgfplotsset{compat=1.18}
\usetikzlibrary{positioning}

\usepackage{graphicx}
\usepackage[caption=false,font=normalsize,labelfont=sf,textfont=sf]{subfig}
\usepackage{booktabs,threeparttable} 
\usepackage{algpseudocode}
\makeatletter
\algrenewcommand\alglinenumber[1]{\footnotesize #1\hspace*{.3em}} % 
\usepackage{multirow}
\usepackage{xcolor}
\usepackage{tikz}
\usepackage{amsthm}
\usepackage{pdflscape}
\usepackage{hyperref}
\usepackage{amsthm}

\usepackage{makecell}

\newtheoremstyle{ieeeplain}%
  {3pt}
  {3pt}
  {\itshape}
  {}
  {\bfseries}
  {.}
  { }
  {}

\theoremstyle{ieeeplain}
\newtheorem{definition}{Definition}
\newtheorem{theorem}{Theorem}

\definecolor{myblue}{HTML}{406FC4}
\begin{document}

\title{Privacy-Preserving Local Energy Trading Considering Network Fees}

\author{Eman alqahtani,~\IEEEmembership{Student Member,~IEEE,} Mustafa A. Mustafa, ~\IEEEmembership{ member,~IEEE}
        % <-this % stops a space
\thanks{This work was supported by EPSRC through EnnCore [EP/T026995/1]. }% <-this % stops a space
% E.A is funded by the Saudi Government through King Abdulaziz University. 
\thanks{Eman Alqahtani is with Department of Computer Science,
The University of Manchester, Manchester, M13 9PL, UK (email: eman.alqahtani@postgrad.manchester.ac.uk}

\thanks{Mustafa A. Mustafa is with Department of Computer Science,
The University of Manchester, Manchester, M13 9PL, UK and with COSIC, KU Leuven, 3001 Leuven, Belgium (email: mustafa.mustafa@manchester.ac.uk)} 
}



\maketitle

\begin{abstract}
Driven by the widespread deployment of distributed energy resources, local energy markets (LEMs) have emerged as a promising approach for enabling direct trades among prosumers and consumers to balance intermittent generation and demand locally. However, LEMs involve processing sensitive participant data, which, if not protected, poses privacy risks. At the same time, since electricity is exchanged over the physical power network, market mechanisms should consider physical constraints and network-related costs. Existing work typically addresses these issues separately, either by incorporating grid-related aspects or by providing privacy protection. To address this gap, we propose a privacy-preserving protocol for LEMs, with consideration of network fees that can incite participants to respect physical limits. The protocol is based on a double-auction mechanism adapted from prior work to enable more efficient application of our privacy-preserving approach. To protect participants’ data, we use secure multiparty computation. In addition, Schnorr’s identification protocol is employed with multiparty verification to ensure authenticated participation without compromising privacy. We further optimise the protocol to reduce communication and round complexity. We prove that the protocol meets its security requirements and show through experimentation its feasibility at a typical LEM scale: a market with 5,000 participants can be cleared in 4.17 minutes.

\end{abstract}


\section{Introduction}
% \subsection{Background and Motivation}
Renewable energy sources, such as photovoltaic panels and wind turbines, are increasingly integrated into power grids. In response, local energy markets (LEMs) have emerged to help coordinate the production of these intermittent resources by incentivising prosumers to proactively trade their surplus energy with neighboring consumers, thereby balancing local supply and demand~\cite{Timothy}. Such markets provide prosumers with several advantages, including autonomy, the ability to express individual preferences~\cite{Morstyn2018} and to trade at competitive prices compared to fixed Feed-in Tariffs.

However, actual electricity exchange occurs over the physical power network, which imposes hard constraints on energy trading. Neglecting these constraints during market mechanism design may lead to constraint violations and compromise the secure operation of the network in practice~\cite{Guerrero}. In parallel, each energy trade within a LEM incurs network-related costs. These include network operation and maintenance costs that the grid operator typically recovers through network usage charges, as well as the costs of compensating for power losses to maintain grid balance. Since LEMs allow participants to make strategic trading decisions, network-related fees can be integrated into the market mechanism to enable participants to efficiently select peers in a way that reduces the network costs imposed on them.

On the other hand, privacy and security concerns remain significant challenges in LEMs. These markets inherently require processing sensitive participant data, such as identities, bid volumes, and prices, to clear the market and allocate traded energy. Disclosing such data directly compromises participants’ privacy, given its correlation with consumption profiles that have been shown to pose privacy risks for individuals~\cite{Quinn}. For instance, malicious entities can analyse such data to infer participants’ behavioral patterns and personal details, such as household presence, sleep schedules, or the activation of specific appliances~\cite{Quinn,Jawurek}. Therefore, collecting and processing unprotected LEM data explicitly violate data protection regulations~\cite{GDPR_2016}, which may, in turn, generate opposition to LEM implementation or limit participation. Besides privacy concerns, the absence of an authentication mechanism for market participants can be exploited to manipulate the market or launch denial-of-service attacks and endanger grid stability~\cite{Dedrick}. Hence, designing a privacy-preserving LEM that incorporates an authentication mechanism while considering the grid aspects is the main motivation behind this work.

% \subsection{Related Work}\label{sec:related_work}
Several works have proposed various LEM mechanism designs that consider physical constraints and network-related fees (based on optimization, auctions, or game-theory). Many of these follow an exogenous approach to avoid the intense involvement of the grid operator in clearing the market~\cite{Guerrero,Kim2020,Khorasany2017,Baroche,Paudel,Khorasany2021}. These approaches can generally be divided into two categories based on when constraint violations are checked and network charges are determined: before or after participant matching. In the latter, after each bid-offer pair is matched, the grid operator validates the feasibility of the transaction on the grid, blocks it if infeasible, or allocates network charges if approved. This approach is followed in~\cite{Guerrero,Kim2020}, where the impact of each transaction on the grid is assessed based on voltage sensitivity analysis. By contrast, in the former approach, bids are initially submitted, and the grid operator then shares the network charges for each potential transaction with the corresponding participants to encourage matching between closer participants, thereby limiting potential adverse impacts of LEMs on the grid. In these designs, the physical constraints are addressed indirectly through network charges, which can be identified based on the electrical distance between participants~\cite{Khorasany2017,Baroche,Paudel}, zone correspondence~\cite{Baroche} or power loss~\cite{Khorasany2021}. However, the reliance on the grid operator in the above works during every market execution to integrate the grid-related aspects, increases the computation complexity and limits the market scalability. In addition, the privacy of participants is not protected in all of the aforementioned works.

Meanwhile, there has been growing interest in developing solutions that address privacy concerns in LEMs~\cite{Dorri,Aitzhan,Aron,Li3,Zhai,Sarenche,Deng,Abidin,Fairouz,Xia,Kamil2024,Xie,Mu,Ping}. These solutions typically use one of the following techniques: anonymisation, differential privacy (DP), homomorphic encryption (HE), or multi-party computation (MPC). In~\cite{Dorri,Aitzhan}, blockchain-based solutions for LEMs are proposed, where each participant uses a new key for every transaction to protect their anonymity. However, despite using different keys, it has been shown that blockchain transactions are still linkable~\cite{Meiklejohn} and that de-anonymisation of metering data is feasible~\cite{Jawurek2}. DP provides stronger privacy guarantees than anonymisation by adding sufficient noise to participants’ data to make them indistinguishable. In~\cite{Li3}, an auctioneer uses the Gaussian mechanism to perturb participants’ valuations before clearing the market and then employs the Exponential mechanism to randomize the allocation process (also used in~\cite{Zhai} for energy allocation) in order to prevent inference of participants’ bids from the auction results. However, DP reduces the accuracy of participants’ data and the economic efficiency of the markets. Moreover, these solutions fully expose participants’ data to the auctioneer.

To preserve privacy without relying on a trusted party having access to participants’ sensitive data, while maintaining data accuracy, another body of work uses HE and/or MPC. HE refers to the schemes that enable computation on encrypted data. Privacy-preserving LEMs based on HE mostly use partially HE, in particular, the Paillier cryptosystem, for private energy aggregation~\cite{Deng,Xie} or market execution~\cite{Sarenche,Xia,Kamil2024}. However, HE operations are computationally intensive, especially on participants’ resource-constrained devices such as smart meters, due to the heavy cryptographic operations involved.  MPC, on the other hand, allows a set of parties to jointly compute a function on private inputs while keeping those inputs secret. In the context of LEMs, MPC-based solutions use secret-sharing protocols to privately aggregate market data (e.g., bid volumes and power constraints)~\cite{Fairouz,Ping}, allocate energy~\cite{Fairouz}, and clear the market~\cite{Abidin}, or garbled circuits to privately compare market variables such as prices~\cite{Xie,Mu}. Although the computational overhead of MPC-based solutions can be relatively low, their primary bottleneck is typically the communication and round complexity associated with non-linear operations. Besides, in all of the aforementioned privacy-preserving solutions, a proper participant authentication mechanism has been mostly neglected and the network-related fees have not been considered. An overview of related work is shown in Table~\ref{tab:related_work}.

% \subsection{Contribution}\label{sec:Contribution}
In summary, previous work has mostly focused either on integrating grid-related aspects or on protecting participants’ privacy and meeting security requirements. To the best of our knowledge, this is the first privacy-preserving solution for LEMs that considers the network fees. Specifically, the main contributions of this paper are as follows:

\begin{itemize}

    \item We design a double auction mechanism for LEMs that accounts for network fees and could incite beneficial trading behavior for the power system. The mechanism is a modification of previous work to reduce computational complexity and improve scalability by eliminating reliance on the grid operator during market execution and allowing participants to select peers beforehand. This allows us to apply our privacy-preserving approach more efficiently.  
     \item Building on the designed mechanism, we propose PNF-LEM, a privacy-preserving LEM protocol based on MPC that protects participants’ data during market execution, including identities, bid volumes, prices, and selected peers. PNF-LEM also authenticates participants while maintaining their privacy using Schnorr's identification protocol with MPC-based verification.
    \item We optimise PNF-LEM to reduce communication and round complexities, thereby improving overall market runtime. In addition, efficient MPC sub-protocols are carefully selected for the instantiation of PNF-LEM.
    \item We analyse the security and complexity of PNF-LEM. We further implement it and evaluate its performance in realistic MPC settings, where the MPC servers are distributed across different European countries. The results show the feasibility of PNF-LEM, as a market with 5{,}000 participants can be cleared within 4.17 minutes.
    
    % \footnote{Our implementation is available at \url{https://github.com/EmanQh/PNF-LEM}}
\end{itemize}

\begin{table}[t]
    \centering     
    \scriptsize
    \caption{Overview of related work.} 
    \label{tab:related_work}
    \begin{tabular}{llcclcc}
        \toprule
         \textbf{Work}& \textbf{\shortstack{Design \\Approach}}  & \textbf{\shortstack{Net. \\Fees}} &  \textbf{Privacy} & \textbf{\shortstack{Privacy\\Technique}} & \textbf{Auth.} & \textbf{\shortstack{No \\TTP}}   \\ 

         \midrule         
        \cite{Paudel,Baroche} & Optimisation & \cmark & \xmark & N/A & \xmark &\xmark\\ 

         \cite{Mu} & Optimisation  & \xmark & \cmark & MPC &\xmark &  \cmark\\

         \cite{Ping} & Optimisation & \xmark & \cmark & MPC & \xmark &  \cmark\\
  
         \cite{Khorasany2021}& Bilateral deal & \cmark & \xmark & N/A & \xmark & \xmark\\ 

         \cite{Dorri,Aitzhan}& Bilateral deal & \xmark& \cmark & Anon.  & \cmark &  \cmark\\ 
         
         \cite{Aron} & Bilateral deal  & \xmark & \cmark & Anon.  & \cmark &  \cmark\\

         \cite{Li3,Zhai} & Double auction & \xmark & \cmark & DP & \xmark &  \xmark\\
         
         \cite{Sarenche} & Double auction & \xmark & \cmark & HE \& Anon. & \cmark &  \xmark\\
         \cite{Guerrero, Khorasany2017}& Double auction &  \cmark & \xmark & N/A & \xmark &  \xmark\\
         \cite{Deng} & Double auction & \xmark & \cmark & HE & \xmark & \xmark\\

        \cite{Abidin,Fairouz} & Double auction &  \xmark & \cmark & MPC & \xmark &  \cmark\\

         \cite{Xia}& Game theory & \xmark & \cmark & HE & \xmark &  \cmark\\  

         \cite{Kamil2024} & Game theory & \xmark & \cmark & HE & \xmark &  \cmark  \\

         \cite{Xie}& Game theory & \xmark & \cmark & HE \& MPC & \xmark &  \cmark\\    

       \midrule
         \textbf{Ours} & Double auction & \cmark & \cmark & MPC &\cmark &  \cmark\\
         
         \bottomrule
    \end{tabular}
\end{table}

The remainder of the paper is organised as follows. Section~\ref{sec:Preliminaries} introduces the preliminaries. Section~\ref{sec:building-blocks} explains the utilised building blocks. Section~\ref{sec:protocol} describes PNF-LEM. Section~\ref{sec:analysis} provides security analysis, while Section~\ref{sec:evaluation} evaluates the performance of PNF-LEM. Finally, Section~\ref{sec:conclusion} concludes.

\section{Preliminaries}\label{sec:Preliminaries}
\subsection{Network Service Pricing}
Various methods have been proposed in the literature to determine how network service charges are computed in LEMs~\cite{Kim}. One way is to define network fees based on the electrical distances between nodes in the network. For instance, the power transfer distribution factor (PTDF) measures the fraction of energy transacted between two nodes that flows through a given line and can be used, as in~\cite{Paudel,Baroche,Khorasany2017}, to define electrical distances. Similar to~\cite{Baroche}, we compute the distance between nodes $i$ and $j$ for a unit power transfer as

\begin{align}
d_{ij} = \sum_{(x,y)\in {\mathcal{L}}} \phi_{ij,xy}
\end{align}

where $\phi_{ij,xy}$ denotes the PTDF of the line from node $x$ to node $y$ for a power injection at node $i$ and a withdrawal at node $j$. The distance-based fee for power transfer from node $i$ to node $j$ can then be computed as

\begin{align}\label{f}
f_{i j} = \frac{u \cdot d_{ij}}{2}
\end{align}

where $u$ is the unit fee that can be set by the distribution system operator (DSO).
The corresponding concrete network charge is given by 
\begin{align}\label{nc}
nc_{ij} = v_{ij} \cdot f_{ij}
\end{align}
where $v_{ij}$ is the amount of energy transferred between nodes $i$ and $j$.

\begin{figure}[t]
\centering
\includegraphics[width=\columnwidth,trim=4 4 4 4,clip]{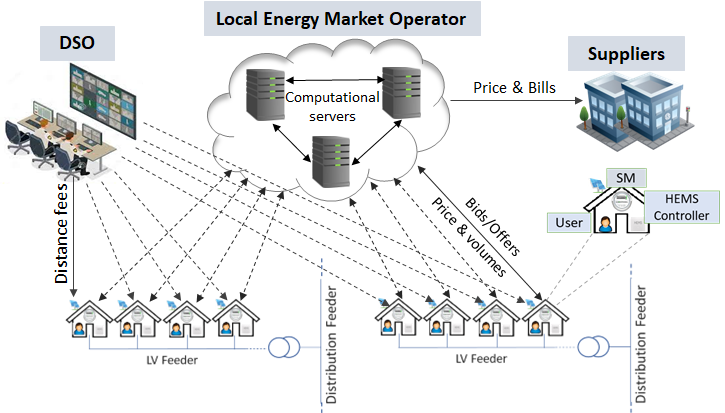}
  \captionsetup{justification=centering}
  \begin{tikzpicture}[baseline={(0,-0.5ex)}]
    % First arrow (Data flow) — moved up to y = 0.2
    \draw[->, thin] (0,0.2) -- (0.5,0.2);
    \node[right] at (0.5,0.2) {\scriptsize Data flow};
    
    % Second arrow (Power flow) — moved up to y = 0.2
    \draw[thin, myblue] (1.8,0.2) -- (2.3,0.2);
    \node[right] at (2.3,0.2) {\scriptsize Power Flow};
\end{tikzpicture}

\caption{System model} 
% \caption{System model (\protect\tikz[baseline={(0,-0.5ex)}]{\draw[->, thin] (0,0)--(0.5,0);} Data flow \protect\tikz[baseline={(0,-0.5ex)}]{\draw[thin, myblue] (0,0)--(0.5,0);} Power Flow).} 

\label{Fig:SystemModel}
\end{figure}
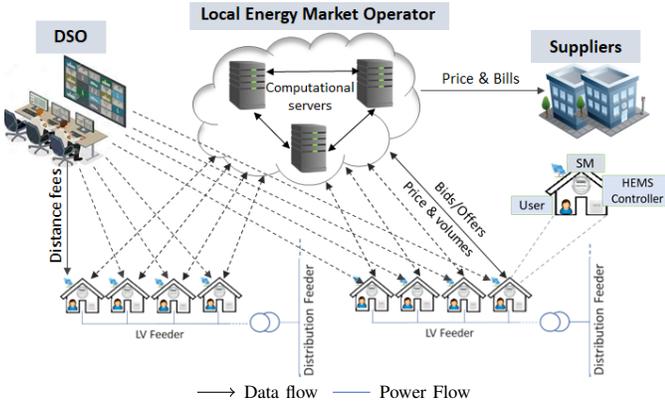

\subsection{System Model}
As shown in Fig.~\ref{Fig:SystemModel}, our system model consists of the following entities. \textit{Users} (either prosumers or consumers) are supported by smart meters (SMs), a home energy management system (HEMS) controller (to which the SMs are connected), and potentially renewable energy sources. Users participate in a LEM through their HEMS controllers via an automated bid submission process configured by the users. The \textit{Local Energy Market Operator (LEMO)} is a complex of three computational servers that clears the LEM, as in Section~\ref{Market-Mechanism}, then generate users’ bills using the identified trading price (TP), accepted bid volumes, and associated network fees. \textit{Suppliers} charge their customers (i.e., users in the LEM) based on these bills. The \textit{DSO} manages and maintains the distribution network within a specific area. It calculates the electrical distance fees for transactions between each pair of users in the LEM area. 

\subsection{Threat Model and Assumptions}
Users and suppliers are assumed to be malicious. Users may attempt to infer other users’ sensitive data, such as unique identifiers, bid volumes, and prices for a given trading period. They may also try to manipulate their own or others’ data, for example, by altering the distance fees between themselves and other peers to reduce network charges, or impersonate other users to influence market outcomes (e.g., energy allocation or TP) to their advantage. Suppliers may also attempt to learn or modify users’ bids/offers to influence TP. External entities are also considered malicious and may eavesdrop on or alter transmitted data to disrupt the market or grid stability. For the LEMO servers, we assume an honest-majority setting in which the adversary may corrupt at most one of three servers. The corrupted server is considered honest-but-curious (a.k.a. semi-honest), meaning it follows the protocol honestly but may attempt to infer sensitive individual data.

We also make the following assumptions. Each LEMO server is controlled by a different stakeholder to avoid collusion. The communication channels are private and authenticated via TLS. 
% All entities are time-synchronised. 
% SMs are sealed and tamper-proof, where any tampering attempts will be readily detected. 
The DSO has a certified public/private key pair, and all entities know its certificate. The DSO shares a distinct symmetric key with each of the LEMO servers. HEMS controllers send their data to LEMO servers through an anonymous communication channel such as Tor~\cite{torproject}.  

\subsection{Security and Privacy Requirements}
 \begin{itemize}     
    \item Privacy preservation: users' unique identifiers, bid volumes, and prices for each trading period should be hidden during market execution from all parties. 
    \item DSO issued tuple authentication: the authenticity of pair transaction fee tuples submitted by users to LEMO should be verified to ensure they were issued by the DSO. 
     \item User authentication: before participating in the LEM, each user should be authenticated and verified as the legitimate owner of the identifier for which their pair transaction fee tuples submitted to LEMO were issued. 
\end{itemize}

\section{Building Blocks}\label{sec:building-blocks}
This section briefly reviews the primitives used in PNF-LEM, namely, MPC and Schnorr’s identification protocol.

\subsection{Multiparty Computation}\label{Sec:MPC}
MPC enables a group of parties to jointly compute a function on private inputs, revealing only the computation result and keeping those inputs secret~\cite{Lindell}. The complexity of MPC protocols is typically evaluated in terms of communication complexity, which refers to the total amount of data exchanged between the parties, and round complexity, which refers to the number of sequential invocations of an MPC primitive during which communication is required. In PNF-LEM, the computation is performed by the three LEMO servers, denoted as $S_1$, $S_2$, and $S_3$. The following presents the underlying MPC primitives and sub-protocols used in this work.

\textbf{Replicated Secret Sharing (RSS).}
PNF-LEM relies on replicated secret sharing~\cite{Ito1989} for the MPC construction. In this scheme, a secret $x$ is shared by selecting three random elements $x_1, x_2, x_3$ such that $x = x_1 + x_2 + x_3$. $S_1$ then holds the share $(x_1, x_2)$, $S_2$ holds $(x_2, x_3)$, and $S_3$ holds $(x_3, x_1)$. We use $[x]$ to denote that $x$ is secretly shared. Addition and multiplication by a public constant can be performed locally by each server. In particular, to calculate $a \cdot [x] + [y]$, each server locally computes $(a \cdot x_i + y_i, a \cdot x_{i+1} + y_{i+1})$, where $i+1$ refers to the index of subsequent server.

\textbf{Multiplication.}
In contrast to addition, multiplying two secretly shared numbers requires a round of communication. In this work, we use the optimised multiplication protocol from~\cite{Araki}, based on RSS, designed for a 3-party setting. The protocol is secure in the presence of semi-honest adversary with at most one corrupted party. It has minimal communication cost, where each server sends only one element per multiplication, by utilising pseudo-random zero sharing~\cite{Ronald}.

\textbf{Comparison.} % nonlinear computation
Several protocols have addressed the problem of performing comparisons in MPC. In this work, we utilise the mixed-circuit technique of \cite{Escudero} with the comparison and equality test protocols of \cite{Catrina}. Escudero et al. \cite{Escudero} improves the efficiency of \cite{Catrina} by enabling transitions between the arithmetic and binary domains, such that linear functionalities are performed in the arithmetic domain, while nonlinear functionalities are carried out in the binary domain. This is achieved using pregenerated extended doubly authenticated bits (edaBits): shared integers in the arithmetic domain whose bit decompositions are shared in the binary domain. The comparison protocol runs in $\mathcal{O}(\log l)$ rounds with $\mathcal{O}(l)$ communication complexity, where $l$ denotes the bit length of the compared values, while the equality test runs in constant rounds with $O(\log l)$ communication complexity.

\textbf{Shuffling.} ~\label{Shuffling}
One of the MPC sub-protocols used in this work is an efficient shuffling protocol presented in Section~3.2 of~\cite{Asharov}, based on RSS. The protocol applies a random permutation $\pi$ to a shared vector $[\mathbf{a}]$ while no server knows $\pi$. This is achieved by applying a composition of three random permutations, $\pi([a]) = \pi_3 \circ \pi_2 \circ \pi_1([a])$, where each $\pi_i$ is known to only two servers. After each two servers apply $\pi_i$ to $[\mathbf{a}]$, they reshare the resulting vector (i.e., generate fresh shares of each vector element) with the third server, so that it cannot learn $\pi_i$ and thereby $\pi$. The protocol runs in three rounds, one per resharing, and incurs a total communication cost of $6m$ elements, where $m$ is the number of vector values.

\textbf{Sorting.}  ~\label{sorting}
Several sorting protocols using MPC have been proposed in the literature, including efficient oblivious radix sorting protocols~\cite{Hamada,Asharov}. Given a key indexed shared vector, these protocols output a sorted shared vector according to the key elements. The protocols work on bitwise shared keys, for which the parties first run a bit-decomposition protocol as in~\cite{Damgard}. For each bit of the keys, the parties securely generate a shared destination vector, which specifies the destination position of each key element after sorting~\cite{Hamada,Asharov}. Both the input vector and the destination vector are then shuffled, so that it is secure to reveal the destination vector and sort the input vector accordingly~\cite{Hamada}. The process is then repeated for each bit of the keys. In this work we use the protocol of~\cite{Hamada}, which runs in $\mathcal{O}(l)$ rounds and exhibits a communication complexity of $\mathcal{O}(l^2 \log l m + l^2 m + l m \log m)$, given a constant number of parties, where $m$ is the number of input vector values.

\textbf{MPC over elliptic curves.} 
PNF-LEM involves performing MPC in the context of elliptic curve calculations. Previous works have shown that MPC protocols over a field based on linear secret sharing can be seamlessly transformed into protocols over elliptic curves~\cite{Smart,Dalskov}. This enables the servers to add secret-shared elliptic curve points or multiply a public field scalar by a secret-shared point locally, as well as to perform non-linear operations, namely to multiply a secret shared scalar by a secret shared point. 

\subsection{Schnorr's Identification Protocol}\label{Sec:Schnorr}
The Schnorr identification protocol is designed based on the discrete logarithm problem to allow the owner of a public key to prove knowledge of the underlying secret key without revealing it~\cite{Schnorr1991}. For an implementation over elliptic curve, let $\mathcal{G} \subseteq E(K)$ be a subgroup of prime order $p$ of an elliptic curve defined over a finite field $K$ and let $G$ be its generator. In the setup phase, the prover selects a private key $d \in \mathbb{F}_p^*$ and generates the corresponding public key as $P = d \cdot G$. The protocol then works as follows in three phases: 

\begin{enumerate}
    \item The prover selects a uniformly random number~$k \in \mathbb{F}_p^*$, computes $R= k \cdot G$, and sends $R$ to the verifier.
    \item The verifier selects a uniformly random challenge $e \in \mathbb{F}_p^*$, and sends it to the prover. 
    \item The prover computes and sends $s = k + e \cdot d$ to the verifier.  
\end{enumerate}
The verifier then accepts the proof if $s \cdot G = R + e \cdot P$. The protocol is known to be suitable for provers of limited computational capability, particularly given that step (1) can be pre-computed, reducing the online computation to just one modular multiplication. Moreover, the protocol can be extended to support batch verification, as recently implemented in bitcoin to speed up signature verification~\cite{bip340}. Specifically, given \( t \) proofs \((R_i, s_i)\) for public keys \(P_i\) and challenges \(e_i\), the verifier selects random scalars \(a_i \in  \mathbb{F}_p^*\) and accepts the proof if $\left( \sum_{i=1}^t a_i s_i \right) \cdot G = \sum_{i=1}^t a_i R_i + \sum_{i=1}^t a_i e_i P_i.$

\textbf{Notations.} We denote an elliptic curve point $X \in \mathcal{G}$ held in secret shared form as $\langle X \rangle$. Secret shared values denoted as $[x]$ are elements of $\mathbb{F}_p$, where $p$ is a large prime that is two bits longer than the input length to enable efficient comparison by computing the difference and extracting the most significant bit as in~\cite{Catrina}. If fixed-point precision is required, the numbers can be scaled by a sufficiently large constant before being secret shared, so they can be represented as elements of $\mathbb{F}_p$. Additionally, we denote by 
$[c] \leftarrow \mathsf{LT}([x],[y])$ and $[c] \leftarrow \mathsf{Eq}([x],[y])$ the execution of the secure comparison (i.e., $[c] \leftarrow ([x]<[y])?1:0$) and equality test (i.e., $[c] \leftarrow ([x]  \leftarrow  [y])?1:0$) protocols, respectively; by $[V'] \leftarrow \mathsf{Shuffle}([V], k)$ and $[V'] \leftarrow \mathsf{Sort}([V], k,o)$ the execution of the secure shuffling and sorting protocols, respectively, 
where $o \in \{0,1\}$ indicates the sorting order ($1$ for ascending and $0$ for descending) and $k$ is the key index; 
by $[x]  \leftarrow  \mathsf{Rand}()$ the generation of random secret shares without interaction using~\cite{Ronald}; and 
by $x = \mathsf{Open}([x])$ the reconstruction of a value from its secret shares.

\subsection{Additional Cryptographic Functionalities}
PNF-LEM also uses the following functionalities: 
\begin{itemize} 
    \item $\sigma \leftarrow \mathsf{Sign(}sk,m)$ and true/false $ \leftarrow \mathsf{Verify}(pk,m,\sigma)$ are public key operations for signing and verification, respectively, using RSA-PSS as defined in~\cite{Moriarty}. 
    % They can be instantiated using RSA-PSS as defined in~\cite{Moriarty}. 
    \item $c \leftarrow \mathsf{Enc}(k,m)$ and $ m \leftarrow \mathsf{Dec}(k,c)$ are symmetric key encryption and decryption functions using AES. 
\end{itemize}

\begin{table}[t]
  \centering
  \footnotesize
  \caption{Notation}
\label{Notations}
\begin{tabular}{p{0.17 \columnwidth} p{0.73 \columnwidth}} 
\toprule
    \textbf{Symbol} & \textbf{Description} \\
       \midrule
       $[ X ]$/$\langle X \rangle$ & Secret shared finite filed element/elliptic curve point \\
       $ID_i$, $sID_u$, $pID_{i,j}$ & Unique identifiers of user~$i$, supplier~$u$, and peer~$j$ selected by user~$i$ \\
        $d_{ij}$ & Electrical transfer distance between user~$i$ and user~$j$ \\ 
        % $f_{ij}$ & Network unit fee for transferring energy from user~$i$ to user~$j$. \\ 
        $t_i$, $v_i$, $p_i$ & Participation type (1 for producer, 0 for consumer), bid volume, and price of user~$i$ \\ 
        $p^*$ & Market clearance price\\
        $a_{ij}, f_{ij}$  & Accepted trading volume, and network unit fee for energy transfer between user~$i$ and user~$j$ \\
        $S^\ell$ & The $\ell$th server for $\ell \in \{1,2,3\}$\\
        $k^\ell$ & A symmetric key shared between DSO and $S^\ell$. \\
        $pk^{d}/sk^{d}$ & Public/private key pair of DSO\\
         $\mathcal{P}_i, \mathcal{P}_i'$ & A set of existing users in the LEM excluding user $i$, a selected subsets of $\mathcal{P}_i$ \\
        $T_{ij}$ & Pair transaction fee tuple, i.e., $(ID_i^\ell, pID_{i,j}, f_{ij}^\ell)$\\
        $B_i$ & Bid/offer details, i.e., $( ID_i, sID_u, v_i, p_i, \{T_{ij}\}_{j \in \mathcal{P}_i'} )$\\
        $C_j^\ell$ & Ciphertext of $\langle pID_{i,j} \rangle$ with $k^\ell$ using  $\mathsf{Enc}$.\\      
        $\sigma_{ij}$ & Signature of $(\langle ID_i^\ell \rangle, C_j^\ell, [f_{ij}^\ell])$ with $sk^{d}$\\   
        $d_i,R_i,e_i,s_i$ & User $i$’s Schnorr secret key, commitment, challenge, and response   \\
        $A_i$ &  Accepted volumes, i.e., $ (ID_i , sID_u , {(a_{ij}, f_{ij})}_{j \in \mathcal{P}_i'})$  \\
        $\mathbf{B}^b/\mathbf{B}^s$ & Vector of sellers’ offers/buyers’ bids tuples\\
        $\mathbf{M}^b/\mathbf{M}^s$ & Selected peer index mappings for sellers/buyers\\
        $\mathbf{A}^b/\mathbf{A}^s$ & Vector of accepted volumes tuples for sellers/buyers\\
         $\mathbf{TA}^b/\mathbf{TA}^s$ & Vector of total accepted volumes for sellers/buyers\\
\bottomrule
\end{tabular}
\end{table}
\section{Privacy-Preserving Local Energy Market}\label{sec:protocol}
In this section, we introduce PNF-LEM, our privacy-preserving protocol for LEMs. We initially present a double-auction market mechanism accounting for network fees such that it incites trading between closer nodes. The mechanism is adapted from prior work to enable more efficient privacy-protection. We then extend this mechanism to PNF-LEM.

\subsection{LEM Mechanism Considering Network Fees} \label{Market-Mechanism}
We present below the stages of our market mechanism incorporating network fees such that it: (i) may limit the potential negative impacts of LEMs on the grid by encouraging market participants to trade energy with their closest partners, and (ii) enables the DSO to recover network usage costs.
\subsubsection{Network service fee computation} \label{stage-1}
Whenever the HEMS controller is installed for a new user~$i$, the DSO performs a one-time computation of the network-related fee as in~(\ref{f}), $f_{ij}$, for all potential transactions between user~$i$ and each user~$j$, $\forall j \in \mathcal{P}_i$, where $\mathcal{P}_i$ is the set of existing users in the LEM excluding user~$i$. For each pair $(i, j)$, the DSO generates and shares two pair transaction fee tuples, $T_{ij} = (ID_i, pID_j, f_{ij})$ and $T_{ji} = ( ID_j, pID_i, f_{ji} )$, with users~$i$ and~$j$, respectively, where $ID$ is the recipient’s identifier, $pID$ is the identifier of the associated peer, and $f$ is the corresponding network fee.

For each trading period, the following stages are executed.  

\subsubsection{Bid submission} \label{stage-2}
The HEMS controller of every user~$i$ selects a subset of their peers, $\mathcal{P}_i' \subseteq \mathcal{P}_i$, based on their preference for lower network fees. They then submit a bid/offer to the LEMO as the tuple $B_i = ( ID_i, sID_u, v_i, p_i, \{T_{ij}\}_{j \in \mathcal{P}_i'} )$ and a participation type $t_i$ (seller or buyer), where $ID_i$ denotes the user's identifier, $sID_u$ the supplier identifier, $v_i$ the amount of energy the user intends to sell/buy, and $p_i$ the offered price.

\subsubsection{Market clearance}  \label{stage-3}
The LEMO executes a double auction mechanism with average pricing to determine the TP, energy allocations, and associated transaction fees. The TP is computed as the average of all submitted bid and offer prices. Energy is then allocated in a greedy manner while respecting users’ peer selections. Assuming indices $i$ and $k$ are used for sellers and buyers, respectively, sellers are sorted in ascending order of $p_i$, and buyers in descending order of $p_k$. For each seller-buyer pair $(i, k)$, energy is allocated as the minimum of $v_i$ and  $v_k$ if the seller and the buyer have mutually selected each other in their respective subsets, $\mathcal{P}_i'$ and $\mathcal{P}_k'$. The outcome of the energy allocation process for each user~$i$ is an accepted trading tuple,
$A_i = ( ID_i, sID_u, \{(a_{ij}, f_{ij})\}_{j \in \mathcal{P}_i'} )$, to be used for billing, where $a_{ij}$ denotes the accepted volume for trading with peer~$j$, and a derived total volume tuple,
$TA_i = (ID_i ,{\sum_{j \in \mathcal{P}_i'} a_{ij}})$, to be provided to the user.

\subsubsection{Informing users and suppliers}
LEMO publishes the trading price (TP) to all users and suppliers. It then sends each user~$i$'s total accepted volume, $\sum_{j \in \mathcal{P}_i'} a_{ij}$, to their HEMS controller. Later, LEMO can compute each user’s bill using $A_i$ and sends the resulting bill to the corresponding supplier.

We note that we use an auction mechanism similar to the one in~\cite{Khorasany2017}, while incorporating network fees, computed as in~\cite{Baroche}, into the energy allocation process. However, in the work of~\cite{Khorasany2017}, estimated network charges are dynamically calculated by the DSO based on the bid volumes received from users in each trading period, which are then used to guide energy allocation. This reliance on a central authority in every trading period can limit scalability, particularly when security and privacy-preserving techniques are applied. In contrast, our approach allows the DSO to predefine fixed fees based on a unit power amount, which can be shared with users once in advance and subsequently used by the users across multiple trading periods. Moreover, by allowing users to preselect a subset of preferred peers and restricting the allocation process to this smaller peer set, we reduce computational complexity (especially for the subsequent privacy-preserving computation) compared to evaluating all possible peers for every user. Final network charges can then be computed during billing, as in~(\ref{nc}), based on the accepted trading volumes. 

\subsection{PNF-LEM}
This section provides a detailed description of PNF-LEM, which builds upon the market mechanism introduced in Section~\ref{Market-Mechanism}. We begin by discussing the prerequisite steps that must be performed once for each user before they can participate in the LEM. We then describe the three main stages of PNF-LEM, executed in every trading period: bid submission, market clearance, and informing users and suppliers.

\subsubsection{Prerequisite}
This section explains the prerequisite steps to enable a user to participate in the LEM, including establishing the user identifier and obtaining signed secret-shared transaction fee tuples from the DSO. In detail, upon installation and integration of the HEMS controller with the SM of user~$i$, the  controller generates a Schnorr private key~$d_i \in \mathbb{F}_p^*$ and computes the corresponding public key, which serves as the user identifier, $ID_i = d_i \cdot G$. The controller then sends the SM identifier and~$ID_i$ to the DSO and the supplier, along with fee-tuples generation request to the latter. Upon receipt of the request, the DSO constructs the transaction fee tuples for user~$i$ for each pair $(i,j)$ (see Section~\ref{stage-1}) as follows: (i) it generates three secret shares of each tuple element, $\langle ID_i^\ell \rangle, \langle pID_j^\ell \rangle$ and $[f_{ij}^\ell]$, one for each server $S_\ell$; (ii) encrypts each peer-identifier share $ \langle pID_j^\ell \rangle$  using the secret key $k^\ell$ shared with $S_\ell$, $C_j^\ell = \mathsf{Enc}(k^\ell, \langle pID_j^\ell \rangle )$, and forms each tuple share $[T_{ij}^\ell] = (\langle ID_i^\ell \rangle, C_j^\ell, [f_{ij}^\ell])$; (iii) signs each tuple share individually, $\sigma^\ell_{ij} = \mathsf{Sign}(sk^{d},[T_{ij}^\ell])$; and (iv) forwards all three signed tuples shares, $\{([T_{ij}^\ell], \sigma^\ell_{ij})\}_{\ell=1}^3$, to the HEMS controller of user~$i$.
$pID_j$ remains hidden from user $i$ because its shares are encrypted with the servers' secret keys, whereas $f_{ij}$ can be reconstructed to enable the user to select peers based on their network fees preferences. To simplify notation, from now on, we omit the superscript~$\ell$ for secret shares.

\subsubsection{Bid Submission}
The HEMS controller of each user~$i$ prepares their bid/offer (see Section~\ref{stage-2}) as follows:  
(i) selects a subset of their transaction fee tuples, $\{([T_{ij}], \sigma_{ij}) \mid j \in \mathcal{P}_i',\ \mathcal{P}_i' \subseteq \mathcal{P}_i\}$;  
(ii) splits each of $ID_i$, $sID_u$, $v_i$, and $p_i$ into three shares, and constructs the bid/offer as $[B_i] = (\langle ID_i \rangle, [sID_u], [v_i], [p_i], \{([T_{ij}], \sigma_{ij})\}_{j \in \mathcal{P}_i'})$;  
(iii) selects a uniformly random number $k_i \in  \mathbb{F}_p^*$, computes the Schnorr commitment $\langle R_i \rangle = [k_i] \cdot G$; and  
(iv) sends $[B_i]$ along with $t_i$ and $\langle R_i \rangle$ to the LEMO servers, each share delivered to its corresponding server, i.e, $S_\ell$. 

\subsubsection{Market clearance}\label{sec:ourProtocol}
This stage has the following steps.

\paragraph{Pre-processing} \label{sec:Preprocess}
Before market execution, the LEMO servers carry out pre-processing steps to perform required verification of authenticity and decrypt the encrypted shares. In detail, upon receipt of $[B_i]$, $t_i$, and $\langle R_i \rangle$ from each user $i$, each $S_\ell$ independently checks the authenticity of every $[T_{ij}]$ by verifying the DSO signature, $\mathsf{Verify}(pk^{d},[T_{ij}], \sigma_{ij}^\ell) \stackrel{?}{=} \text{true}$. Next, the LEMO servers jointly perform distributed verification to ensure that $[B_i]$ was submitted by the legitimate owner of $\langle ID_i \rangle$, which was registered during the installation of HEMS controller. To this end, the LEMO servers engage in an instance of Schnorr's identification protocol with user~$i$ to verify that the user knows the private key $d_i$ corresponding to $ \langle ID_i \rangle$. Specifically, the servers first select a uniformly random challenge $e_i \in \mathbb{F}_p^*$, and forward it to the user. The user then computes $s_i = k_i + e_i \cdot d_i$ and sends $s_i$ to all servers. To improve efficiency, the LEMO servers perform batch verification of all user proofs upon receiving them, as follows:  
(i) select a random number $a_{ij} \in \mathbb{F}_p^*$ for every $[T_{ij}]$;  
(ii) compute $ \langle X \rangle = \sum_{(i,j) \in \mathcal{U}} a_{ij} \langle R_i \rangle +  \sum_{(i,j) \in \mathcal{U}} a_{ij} e_i \langle ID_i \rangle $, where $\mathcal{U}$ denotes the set of all users selected peer pairs; (iii) reveal the result, $X = \mathsf{Open}(\langle X \rangle )$; and (iv) accept the batch of proofs if $\left( \sum_{(i,j) \in \mathcal{U}} a_{ij} s_i \right) \cdot G = X$. Upon successful verification, each $S_\ell$ independently decrypts every $C_j^\ell$ it received as an element of $[T_{ij}]$, $\langle pID_j \rangle = \mathsf{Dec}(k^\ell, C_j^\ell)$.

\paragraph{TP Calculation and Energy Allocation} \label{sec:Allocation}
The LEMO servers execute the auction to determine the TP and the energy allocation in an oblivious fashion. The resulting outcome includes (i) the TP, (ii) two vectors of accepted volumes, $[\mathbf{A}^s]$ for sellers and $[\mathbf{A}^b]$ for buyers, where $[A_i] = (\langle ID_i \rangle, [ sID_u ], {([a_{ij}], [f_{ij}])}_{j \in \mathcal{P}_i'})$, and (iii) two derived vectors, $[\mathbf{TA}^b]$ and $[\mathbf{TA}^b]$, where~$[TA_i] = (\langle ID_i \rangle,{\sum_{j \in \mathcal{P}_i'} [a_{ij}]})$. Initially, bids/offers $[B_i]$ are classified by $t_i$ into a vector of sellers’ offers, $[\mathbf{B}^s]$, and buyers’ bids, $[\mathbf{B}^b]$. For notational simplicity, we use $i$ and $k$ to index seller and buyer elements, respectively, in this phase. The TP is then computed as:  
{
\[
p^* = \frac {\mathsf{Open}\Big(\sum_{i=1}^{|\mathbf{B}^s|} [p_i] + \sum_{k=1}^{|\mathbf{B}^b|} [p_k]\Big)}{|\mathbf{B}^s| + |\mathbf{B}^b|}
\]
}
Next, $[\mathbf{B}^s]$ and $[\mathbf{B}^b]$ are securely sorted in ascending order of $[p_i]$ and descending order of $[p_k]$, respectively, giving
$[\mathbf{B'}^{s}] = \mathsf{Sort}([\mathbf{B}^s], k_{p_i}, 1)$ and
$[\mathbf{B'}^b] = \mathsf{Sort}([\mathbf{B}^b], k_{p_k}, 0)$.
\begin{algorithm}[t]
\small
\caption{Peer Index Mappings} % Tensors
\label{alg:Peer_Mappings1}
\footnotesize
\begin{algorithmic}[1]

\Statex \textbf{Input:} Peer selections $\{(\langle ID \rangle , \langle pID \rangle, [f])\}_{j=1}^{|\mathcal{P}_i'|}$ for each of $|\mathbf{B}^s|$ sellers and $|\mathbf{B}^b|$ buyers.
\Statex \textbf{Output:} Peer mappings for sellers $[\mathbf{M}^s]\! \in\! \{0,1\}^{|\mathbf{B}^s|\!\times\!|\mathcal{P}'|\!\times\!|\mathbf{B}^b|}$, and buyers
$[\mathbf{M}^b]\! \in\! \{0,1\}^{|\mathbf{B}^b|\!\times\!|\mathcal{P}'|\!\times\!|\mathbf{B}^s|}$.

% Peer mappings (binary)
\For{$i = 1$ to $|\mathbf{B}^s|$} \textbf{parallel}
  \For{$j = 1$ to $|\mathcal{P}_i'|$} 
    \For{$k = 1$ to $|\mathbf{B}^b|$} 
        \State $[M]_{i,j,k} \gets \mathsf{Eq}(\langle pID\rangle_{i,j} , \langle ID\rangle_k )$
      \EndFor
  \EndFor
\EndFor

\For{$k = 1$ to $|\mathbf{B}^b|$} \textbf{parallel}
  \For{$j = 1$ to $|\mathcal{P}_k'|$}
      \For{$i = 1$ to $|\mathbf{B}^s|$}  
        \State $[M]_{k,j,i} \gets  \mathsf{Eq}(\langle  pID\rangle_{k,j}, \langle ID\rangle_i)$
      \EndFor
  \EndFor
\EndFor
\end{algorithmic}
\end{algorithm}

\begin{algorithm}[t]
\small
\caption{Energy Allocation} \label{alg:MarketClearance1}
\footnotesize
\begin{algorithmic}[1]
\Statex \textbf{Input:} Sorted bids $[\mathbf{B'}^b]$, offers $[\mathbf{B'}^s]$, and peer mappings $[\mathbf{M}^s]$, $[\mathbf{M}^b]$.
\Statex \textbf{Output:} Accepted volumes for sellers $[\mathbf{A^s}]$, and buyers $[\mathbf{A^b}]$, where $[A_i] = (\langle ID \rangle, [ sID ], \{[a], [f]\}_{j=1}^{|\mathcal{P}_i'|})$.

\State $[\overline{V}] \gets \{0_1,..,0_{|\mathcal{P}'|^2}\}$

\For{$i = 1$ to $|\mathbf{B}^s|$}
  \For{$k = 1$ to $|\mathbf{B}^b| $}
    \State $[c] \gets \mathsf{LT}([v]_k,[v]_i)$  
%    \Comment{Phase 1: Compute all [\overline{v}]_x values in parallel (no race conditions)}
    \For{$j = 1$ to $|\mathcal{P}_i'|$} \textbf{parallel} %\Comment{In parallel}
      \For{$j_2 = 1$ to $|\mathcal{P}_k'|$}
        \State $x \gets (j-1) \cdot |\mathcal{P}_i'| + j_2$
        \State $[\overline{v}]_{x} \gets \left( [c] \cdot ([v]_k - [v]_i) + v_i \right) \cdot [M]_{i,j,k} \cdot [M]_{k,j_2,i} $
      \EndFor
    \EndFor
    \For{$j = 1$ to $|\mathcal{P}_i'|$} \Comment{Sequential} %accumulation
      \For{$j_2 = 1$ to $|\mathcal{P}_k'|$}
        \State $x \gets (j-1) \cdot |\mathcal{P}_i'| + j_2$
        \State $[a]_{i,j} \gets [a]_{i,j} + [\overline{v}]_{x}$
        \State $[a]_{k,j_2} \gets [a]_{k,j_2} + [\overline{v}]_{x}$
        \State $[v]_i \gets [v]_i - [\overline{v}]_{x}$
        \State $[v]_k \gets [v]_k - [\overline{v}]_{x}$
      \EndFor
    \EndFor    
  \EndFor
\EndFor
\end{algorithmic}
\end{algorithm}

\textbf{Na\"{\i}ve energy allocation.} The next step is to obliviously distribute energy from sellers to buyers for those who have mutually selected each other. This can be achieved by sequentially iterating over all seller-buyer pairs $(i,k)$, computing the minimum of $[v_i]$ and $[v_k]$ using secure comparison, then verifying mutual peer selection condition using secure equality test and allocating energy accordingly. However, due to sequential invocations of secure comparisons and equality tests, this solution runs in $\mathcal{O}(|\mathbf{B}^s| |\mathbf{B}^b| (\log l + |\mathcal{P}'|))$ rounds with $\mathcal{O}(|\mathbf{B}^s| |\mathbf{B}^b| (\ell + |\mathcal{P}'| log \ell + |\mathcal{P}'|^2))$ communication complexity.

\textbf{Overhead reduction.} To reduce the overhead, we separate out expensive operations that can be executed in parallel to batch their communication rounds. In particular, the LEMO servers could in advance perform the equality tests for peer selection checks in parallel and store the results for later use in energy allocation, as shown in Algorithm~\ref{alg:Peer_Mappings1}. The algorithm outputs two three-dimensional binary tensors, $[\mathbf{M}^s]\! \in\! \{0,1\}^{|\mathbf{B}^s|\!\times\!|\mathcal{P}'|\!\times\!|\mathbf{B}^b|}$ for sellers and $[\mathbf{M}^b]\! \in\! \{0,1\}^{|\mathbf{B}^b|\!\times\!|\mathcal{P}'|\!\times\!|\mathbf{B}^s|}$ for buyers, to represent their selected peer index mappings. Specifically, $M^s_{i,j,k}=1$ if buyer~$k$ is the $j$-th selected peer by seller~$i$ (i.e., $pID_{i,j} = ID_k$) and $0$ otherwise. The tensor $[\mathbf{M}^b]$ is defined analogously.

After precomputing the peer index mappings $[\mathbf{M}^s]$ and $[\mathbf{M}^b]$, the servers perform the energy allocation and outputs $[\mathbf{A}^s]$ and $[\mathbf{A}^b]$ as shown in Algorithm~\ref{alg:MarketClearance1}. In detail, for each seller-buyer pair $(i,k)$, the algorithm uses $[\mathbf{M}^s]$ and $[\mathbf{M}^b]$ to identify, in parallel, the accepted volumes for each mutual peer selection check (either the the minimum of $[v_i]$ and $[v_k]$ or zero) (\textit{line~8}); stores the results (\textit{lines~14–15}); and decrements the remaining offered volumes (\textit{lines~16–17}). Since Algorithm~\ref{alg:Peer_Mappings1} can be batched in two constant rounds, this reduces the overall round complexity, compared with the naive solution, to $\mathcal{O}(|\mathbf{B}^s| |\mathbf{B}^b| \log \ell)$. However, the communication cost is equally high, as the number of secure equality test and comparison invocations remains the same. 

\textbf{Further overhead reduction.} To reduce overhead further, we observe that peer index mappings can be revealed, $\mathbf{M}^s$ and $\mathbf{M}^b$, provided that the output vectors~$[\mathbf{TA}^s]$~and~$[\mathbf{TA}^b]$ (derived from $[\mathbf{A}^s]$~and~$[\mathbf{A}^b]$)~are shuffled before reconstructing users’ identifiers for results distribution. This will reduce the number of secure comparisons during energy allocation. Moreover, under this relaxation, the mappings $\mathbf{M}^s$ and $\mathbf{M}^b$ can be computed without expensive secure equality tests by masking identifiers and then performing equality tests in the clear, as shown in Algorithm~\ref{alg:Peer_Mappings2}. Note that it is crucial to use scalar multiplication for masking (\textit{Line~6}) to hide not only the identifiers but also their differences, which if leaked, may allow recovery of the identifiers since the set of users’ identifiers is known to the LEMO servers. Energy allocation then proceeds as shown in Algorithm~\ref{alg:MarketClearance2},~where a secure comparison of $[v_i]$ and $[v_k]$ is performed (\textit{Line~9}) only if mutual selection is satisfied, i.e., $M_{i,j,k} = M_{k,j,i} = 1$. By avoiding secure equality tests and reducing comparisons, this solution achieves the complexity of $\mathcal{O}(|\mathbf{B}^s| |\mathcal{P}'| \log \ell)$ rounds and $\mathcal{O}(|\mathbf{B}^s| |\mathcal{P}'| (|\mathbf{B}^b| + \ell))$~communications.

\subsubsection{Informing users and suppliers}\label{sec:ourProtocol}
After the LEM is cleared, the LEMO servers publish $p^*$. To hide bid/offer ordering and peer index mappings before distributing results, the vectors of total accepted volumes are shuffled,~$[\mathbf{TA}^{s'}] \leftarrow \mathsf{Shuffle}([\mathbf{TA}^s], k)$ and
$[\mathbf{TA}^{b'}] \leftarrow \mathsf{Shuffle}([\mathbf{TA}^b], k)$. The servers then reconstruct~$ID_i$ for each~$TA_i'$ and send their shares of~$\sum_{j \in \mathcal{P}_i'} a_{ij}$ to the designated user’s HEMS controller identified by~$ID_i$. Later, the servers can utilise~$[\mathbf{A}^s]$~and~$[\mathbf{A}^b]$ to compute the users' bills, including network charges, and report them to the corresponding suppliers after bills shuffling based on reconstructed~$sID_u$. Note that this work focuses on protecting privacy during market execution; privacy during billing has been addressed in prior work~\cite{Eman2023}.

\section{Security Analysis}\label{sec:analysis}

\textbf{Privacy-Preservation.} We analyse the privacy-preservation requirement of PNF-LEM under the universal composability (UC) framework~\cite{Canetti,Canetti2}. The framework provides guarantees by which a protocol proven secure in this framework is assured to maintain its security when it is run sequentially or concurrently with other protocols.
\begin{algorithm}[t]
\footnotesize
\caption{Peer Index Mappings: ID Mask and Reveal } \label{alg:Peer_Mappings2}
\begin{algorithmic}[1]
\Statex \textbf{Input:} Peer selections, $\{(\langle ID \rangle, \langle pID \rangle, [f])\}_{j=1}^{|\mathcal{P}'|}$ for each of $|\mathbf{B}^s|$ sellers and $|\mathbf{B}^b|$ buyers.
\Statex \textbf{Output:} Peer mappings for sellers $[\mathbf{M}^s]\! \in\! \{0,1\}^{|\mathbf{B}^s|\!\times\!|\mathcal{P}'|\!\times\!|\mathbf{B}^b|}$, and buyers
$[\mathbf{M}^b]\! \in\! \{0,1\}^{|\mathbf{B}^b|\!\times\!|\mathcal{P}'|\!\times\!|\mathbf{B}^s|}$.
\For{$i = 1$ to $|\mathbf{B}^s|$} \textbf{parallel}
  \For{$j = 1$ to $|\mathcal{P}_i'|$}  
       \For{$k = 1$ to $|\mathbf{B}^b|$} 
        \State $[r]_{i,j,k}  \gets \mathsf{rand()}$
        \State $M_{i,j,k} \gets \mathsf{open}([r]_{i,j,k} \cdot ( \langle pID \rangle_{i,j}  - \langle ID \rangle_k  )) \stackrel{?}{=} 0$
      \EndFor
  \EndFor
\EndFor

\For{$k = 0$ to $|\mathbf{B}^b|$} \textbf{parallel}
 \For{$j = 0$ to $|\mathcal{P}_k'|$}
  \For{$i = 0$ to $|\mathbf{B}^s|$}
        \State $[r]_{k,j,i}  \gets \mathsf{rand()}$
        \State $M_{k,j,i} \gets \mathsf{open}([r]_{k,j,i} \cdot ( \langle pID \rangle_{k,j}  - \langle ID \rangle_i  )) \stackrel{?}{=} 0$
      \EndFor
  \EndFor
\EndFor

\end{algorithmic}
\end{algorithm}
\begin{algorithm}[t]
\footnotesize
\caption{Optimised Energy Allocation} \label{alg:MarketClearance2}
\begin{algorithmic}[1]
\Statex \textbf{Input:} Sorted bids $[\mathbf{B'}^b]$, offers $[\mathbf{B'}^s]$, and peer mappings $[\mathbf{M}^s]$, $[\mathbf{M}^b]$.
\Statex \textbf{Output:} Accepted volumes for sellers $[\mathbf{A^s}]$, and buyers $[\mathbf{A^b}]$, where $[A_i] = (\langle ID \rangle, \langle sID \rangle, \{[a], [f]\}_{j=1}^{|\mathcal{P}'|})$.
\For{$i=1$ to $|\mathbf{B}^s|$}
  \For{$k=1$ to $|\mathbf{B}^b|$}
    \State $[\overline V] \gets \{0_1,\dots,0_{|\mathcal{P}'|^2}\}$ 
    \For{$j=1$ to $|\mathcal{P}_i'|$}
      \If{$M_{i,j,k} = 1$}
      \For{$j_2=1$ to $|\mathcal{P}_k'|$}
       \If{$M_{k,j_2,i} = 1$}
        \State $x \gets (j-1) \cdot |\mathcal{P}_i'| + j_2$
        \State $[c] \gets \mathsf{LT}( [v]_k, [v]_i)$
        \State $[\overline{v}]_x \gets \left( [c] \cdot ([v]_k - [v]_i) + v_i \right) $
        \State $[a]_{i,j},[a]_{k,j_2} \gets [\overline{v}]_x$ 
        \State $[v]_i \gets [v]_i - [\overline{v}]_x$
        \State $[v]_k \gets [v]_k - [\overline{v}]_x$
       \EndIf
      \EndFor
     \EndIf
    \EndFor
  \EndFor
\EndFor

\end{algorithmic}
\end{algorithm}

\begin{definition}[UC emulation~\cite{Canetti}] A real protocol $\pi$ UC-realizes an ideal functionality $\mathcal{F}$ if for any real-world adversary $\mathcal{A}$, there exists a simulator $\mathcal{S}$ such that no environment $\mathcal{Z}$ is able to distinguish with a non-negligible probability between an interaction with $\mathcal{A}$ and real parties running protocol $\pi$ and an interaction with $\mathcal{S}$ and dummy parties accessing $\mathcal{F}$, i.e., $REAL_{\pi,\mathcal{A},\mathcal{Z}} \approx IDEAL_{\mathcal{F},\mathcal{S},\mathcal{Z}}$.
\end{definition}

\begin{definition}[Composition Theorem~\cite{Canetti}] \label{Composition}
Let $\rho$, $\varphi$, and $\pi$ be protocols such that $\varphi$ is a subroutine of $\pi$, $\rho$ UC-emulates $\varphi$, and $\pi$ UC-realizes an ideal functionality $\mathcal{F}$, then protocol $\pi^{\varphi \rightarrow \rho}$ UC-realizes $\mathcal{F}$.
\end{definition}

We now prove the security of PNF-LEM based on Algorithms~\ref{alg:Peer_Mappings2} and~\ref{alg:MarketClearance2}. We begin by defining the ideal functionality $\mathcal{F}_{PLEM}$ for privacy-preserving LEM. 

\begin{definition}[$\mathcal{F}_{\text{PLEM}}$]
Upon receiving bid/offer tuples $B_i$ and their corresponding participation types $t_i$ from the user HEMS controllers, the ideal functionality $\mathcal{F}_{\text{PLEM}}$ sorts the vector of bids $B_b'$ and offers $B_s'$ (as described in Section~\ref{stage-3}), computes and sends the peer index mappings $M^s$ and $M^b$ to the adversary, then computes the market outcomes (i.e., the trading price $p^*$ and accepted volumes $A_b$ and $A_s$) and returns the results to the designated parties (i.e., users and suppliers).
\end{definition}
Each of the MPC subprotocols used in PNF-LEM has been proven UC-secure, including multiplication~\cite{Araki}, comparison using edaBits~\cite{Escudero}, shuffling and sorting~\cite[Section~4.2]{Hamada}, elliptic curve-based operations~\cite{Smart}, and pseudorandom secret sharing~\cite[Section~2.2]{Hamada}. According to the Composition Theorem~\ref{Composition}, this guarantees that no environment can distinguish, with a non-negligible probability, between the case in which it observes an instance of PNF-LEM using these subprotocols and an instance where the subprotocols are replaced by the corresponding ideal functionalities: multiplication $\mathcal{F}_{\text{Mult}}$, edaBits $\mathcal{F}_{\text{EdaBits}}$, sorting $\mathcal{F}_{\text{Sort}}$, operations over elliptic curve $\mathcal{F}_{\text{ECC}}$, and pseudorandom secret sharing $\mathcal{F}_{\text{Rand}}$, as defined in~\cite{Araki,Escudero,Hamada,Smart}. We then have the following theorem, proven below:

\begin{theorem}
    PNF-LEM securely emulates $\mathcal{F}_{\text{PLEM}}$ in the ($\mathcal{F}_{\text{Mult}}$,$\mathcal{F}_{\text{EdaBits}}$, $\mathcal{F}_{\text{Sort}}$, $\mathcal{F}_{\text{ECC}}$, $\mathcal{F}_{\text{Rand}}$)-hybrid model.
\end{theorem}

\textit{PROOF.} The environment has no view of the subprotocols used in PNF-LEM, since they are substituted with the invocations of the corresponding ideal functionalities. The remaining computations are either local or involve interactions for opening random elements that can be simulated by the simulator$~\mathcal{S}$. One example is the reveal of the masked difference between identifiers during peer index mappings of Alg.~\ref{alg:Peer_Mappings2} (\textit{line 6}). If the difference is not zero, then the revealed element seen by the environment is random. Based on $M^s$ and $M^b$ received from $\mathcal{F}_{\text{PLEM}}$, $~\mathcal{S}$ can determine when the difference is not zero (i.e., identifiers are unequal). If this is the case, $~\mathcal{S}$ selects a uniformly random number $r\in [1, n-1]$ and sends $r \cdot G$ to the environment as its incoming message; otherwise, it sends~zero. 

\textbf{DSO issued tuple authentication.}
The DSO signs each pair transaction fee tuple share $[T_{ij}]$ it generates 
using RSA-PSS that is existentially unforgeable under chosen-message attacks in the random oracle model, assuming the integer factorisation problem is hard~\cite{Bellare}. Each user then attaches the DSO's signature of each $[T_{ij}]$ they include in $[B_i]$, allowing the LEMO servers to ensure its integrity and DSO authenticity. 

\textbf{User authentication.}
Upon installation of each user's HEMS controller, the DSO registers the user identifier $ID_i$, representing its Schnorr public key, generates every possible $[T_{ij}]$ embedding $\langle ID_i \rangle$, and digitally signs them. For each submitted $[B_i]$, the LEMO servers verify the DSO’s signature on every included $[T_{ij}]$. The user then runs an instance of Schnorr’s protocol with the servers, providing a signature $s_i$ to prove knowledge of the private key corresponding to each $\langle ID_i \rangle$ contained in $[B_i]$. Successful verifications confirm that $ID_i$ was registered by the DSO and that the user is the legitimate owner of $ID_i$, for whom every $[T_{ij}]$ included in $[B_i]$ was issued. This holds since RSA-PSS is provably secure, and Schnorr’s protocol is a secure proof-of-knowledge protocol under the discrete logarithm assumption~\cite{Schnorr1991}.

\section{Performance Evaluation}\label{sec:evaluation}
\subsection{Theoretical Complexity}
We focus our complexity analysis on the market clearance, the most demanding stage executed for every trading period. We include in our analysis the round and the communication complexity. The complexity of the market clearance preprocessing is $\mathcal{O}(N |\mathcal{P}'|)$, where $N = |\mathbf{B}^s|+|\mathbf{B}^b|$. Moreover, during preprocessing, Schnorr’s identification protocol requires two rounds of interaction between each user and the LEMO servers, and one round between the LEMO servers to perform a single $\mathsf{Open}([x])$ during batch verification. The remaining preprocessing operations are local. For energy allocation, the overall complexity of Algorithms~\ref{alg:Peer_Mappings2} and~\ref{alg:MarketClearance2} is $\mathcal{O}(|\mathbf{B}^s| |\mathbf{B}^b| |\mathcal{P}'|^2)$. Recall that they run overall in $\mathcal{O}(|\mathbf{B}^s| |\mathcal{P}'| \log \ell)$ rounds with $\mathcal{O}(|\mathbf{B}^s| |\mathcal{P}'| (|\mathbf{B}^b| + \ell))$ communication complexity. Moreover, sorting offers and bids runs in $\mathcal{O}(\ell)$ rounds and has $\mathcal{O}(\ell^2 \log \ell N + \ell^2 N + \ell (|\mathbf{B}^s| \log |\mathbf{B}^s| +|\mathbf{B}^b| \log |\mathbf{B}^b|))$ communication complexity, while the shuffling can be achieved in $\mathcal{O}(1)$ rounds and $\mathcal{O}(N)$ communication complexity. 

\subsection{Experimentation}
\textbf{Dataset.} We used the realistic dataset generated in~\cite{Abidin} to simulate bid volumes and offered prices over a 30 minute trading period (a typical trading period for LEMs). For peer selection, we assumed that each user selects three peers such that at least one of their selected peers of the opposite type (seller or buyer) has also mutually selected them. We then randomly chose 50\% of the users to have one additional mutual peer selection. We initiated our experiments with 1000 users and gradually scaled to 5000 users. We evenly divided users between producers and consumers.

\textbf{Experiment Setup.} Our experiments run the three LEMO servers and DSO on AWS c5a.8xlarge machines. Each server has a 32-core single-threaded Intel Xeon processor and 64 GB of memory. The servers were located in different regions (Frankfurt, Zurich, Milan), where the maximum RTTs between each pair of servers were 4.87~ms, 7.28~ms, and 10.90~ms. The operations of the HEMS controllers were evaluated on a laptop with an Intel Core i7 CPU and 8 GB of memory.

We implemented PNF-LEM using the MP-SPDZ framework~\cite{Keller2}, as it supports the underlying MPC subprotocols employed in this work. The preprocessing operations (Section~\ref{sec:Preprocess}) were implemented using the low-level C++ interface of MP-SPDZ and the OpenSSL library. For $\mathsf{Enc}$, we used AES in GCM mode with 256-bit keys and for $\mathsf{Sign}$, we used RSA-PSS with 2048-bit keys and SHA-256. Schnorr’s identification protocol was instantiated over the \texttt{secp256k1} curve. TP calculation and energy allocation (Section~\ref{sec:Allocation}, using Algorithms~\ref{alg:Peer_Mappings2} and~\ref{alg:MarketClearance2}) were implemented using MP-SPDZ’s high-level interface based on Python. Our implementation utilises MP-SPDZ’s multi-threading and loop parallelisation for communication-round optimisation, using $\mathsf{for\_range\_opt\_multithread}$, where the number of threads is set to 32 and the optimisation budget to 100{,}000. We tested input values of lengths 32 and 64 bits. The implementation is publicly available at \url{https://github.com/EmanQh/PCP-LEM}. 

\begin{table}
    \centering
     
    \caption{Running time of market clearance preprocessing operations} 
    \label{tab:preprocessing_time}
    \footnotesize
    \renewcommand{\cellgape}{\Gape[4pt]} % Adds a little breathing room
    \begin{tabular}{ccccc}
        \toprule
         % \thead{\textbf{Bids}} & \thead{\textbf{RSA} \\ \textbf{Ver (ms)}} & \thead{\textbf{Schnorr} \\ \textbf{Ver (s)}} & \thead{\textbf{AES} \\ \textbf{Dec (ms)}} & \textbf{Total (s)}  \\ 
            \textbf{Bids} & \textbf{RSA Ver (ms)} & \textbf{Schn. Ver (s)} & \textbf{AES Dec} (ms) & \textbf{Total} \\
         \midrule
        1000 & 70.62 & 3.52 & 16.55 & 3.61 \\
        2000 & 127.99 & 6.64 & 24.19 & 6.79 \\
        3000 & 211.02 & 9.75 & 49.22 & 9.99 \\
        4000 & 281.01 & 12.87 & 65.23 & 13.16 \\
        5000 & 350.84 & 15.99 & 81.01 & 16.43 \\
         \bottomrule
    \end{tabular}
\end{table}

\textbf{Results.} 
Based on our experiments, we present below the computation cost for each stage of PNF-LEM.

\textit{Prerequisite.}
Upon installation of the HEMS controller for user~$i$, it performs a scalar multiplication to generate the user's Schnorr public key, which costs about $0.82~\text{ms}$. Moreover, the DSO performs $6 \times (N-1) \times (\mathsf{Share} + \mathsf{Enc} + \mathsf{Sign})$ operations (recall that for each pair $(i,j)$, we have two tuples of three elements, for each of which three shares are generated), where $\mathsf{Share}$ costs $0.62~\mu\text{s}$, $\mathsf{Enc}$ costs $5.4~\mu\text{s}$, and $\mathsf{Sign}$ costs $0.65~\text{ms}$. For a market size of $5{,}000$ users, our experiments show that upon a new HEMS controller installation, the DSO takes $19.67~\text{s}$ in total for generating shares, encryption, and signing. 

\begin{table}
    \centering   
    \caption{Results of market clearance energy allocation (\textnormal{64-bit})}
    \label{tab:Allocation_time}
    \footnotesize
    \begin{tabular}{ccccccc}
        \toprule
         \multirow{2}{*}{\textbf{Bids}} &  \multicolumn{2}{c}{\textbf{Time (s)}} & \multicolumn{2}{c}{\textbf{Rounds}} & \multirow{2}{*}{\textbf{CPU time (s)}}  
          \\
          \noalign{\vspace{0.5ex}}\cline{2-5}  \noalign{\vspace{0.5ex}} 
            & Online & Offline & Online & Offline &  \\ 
         \midrule
          1000 &  14.31 & 3.33 & 9370 & 22868 & 0.64   \\
           2000 & 55.23 & 10.33 & 59362 & 89748 & 1.69   \\
          3000 & 87.84 & 25.36 & 123862 & 195651 & 5.61 \\
          4000 & 129.78 & 41.17 & 212362  & 350822 & 6.70 \\
           5000 & 173.25 & 67.16 & 324853  & 545600 & 15.22 \\
         \bottomrule
    \end{tabular}
\end{table}

 \begin{figure}[b]
            \centering
            \begin{tikzpicture}[scale=0.61]
                   \begin{axis}[
                ybar stacked,
                bar shift=-4pt,             % Increased left shift
                enlarge x limits=0.25,      % More horizontal space
                ylabel={Time (s)},
                xlabel={Users},
                ymin=0,
                ymax=250,
               ytick={0,100,200}, 
                xtick={1,2,3,4,5},
                xticklabels={1000,2000,3000,4000,5000},
                ymajorgrids=true,
                axis x line*=bottom,
                axis y line*=left,
                clip=false,
            ]
             % Offline Field 3 & 4, 32 bits
            \addplot+[fill=black!30, draw=black!60] coordinates {
            (1,1.94) (2,6.43) (3,14.25)(4,24.15) (5,30.53)};
            % Online Field 3 & 4, 32 bits
            \addplot+[fill=black!70, draw=black] coordinates {
                (1,10.18) (2,48.29) (3,81.21)(4,119.94) (5,166.64) };

            \end{axis}
    
            \begin{axis}[
               ybar stacked,
               bar width=9pt,             
               bar shift=+7pt,             % Increased right shift 
               enlarge x limits=0.25,
               ymin=0,
               ytick={0,100,200}, 
                ymax=250,
                axis x line=none,
                axis y line=none,
                tick style={draw=none},
                xtick={1,2,3,4,5},
                ytick=\empty,
                axis x line*=bottom,
                axis y line*=left,
                clip=false,
            ]
                                            % Offline Field 3 & 4, 64 bits
                \addplot+[fill=gray!50, draw=gray!70] coordinates {
                (1,3.33) (2, 10.33) (3,25.36)(4,41.17) (5,67.16)};
                % Online Field 3 & 4, 64 bits
                \addplot+[fill=gray!80, draw=gray!90] coordinates {
                (1,14.31) (2,55.23) (3,87.84)(4,129.78) (5,173.25)};
               
         \end{axis}
         % Manual legend
        \node[anchor=north west,
              draw=black,          % border
              fill=white,          % background
              line width=0.4pt,    
              font=\scriptsize,    
              inner sep=1pt]       
        at (rel axis cs:0.2,0.98) {
            \begin{tabular}{@{\hskip 3pt}l@{\hskip 3pt}l@{\hskip 3pt}}
                \tikz{\draw[fill=black!30, draw=black!60] (0,0) rectangle (0.15,0.15);} & Offline 32-bit \\
                \tikz{\draw[fill=black!70, draw=black] (0,0) rectangle (0.15,0.15);} & Online 32-bit \\
                \tikz{\draw[fill=gray!50, draw=gray!70] (0,0) rectangle (0.15,0.15);} & Offline 64-bit \\
                \tikz{\draw[fill=gray!80, draw=gray!90] (0,0) rectangle (0.15,0.15);} & Online 64-bit \\
            \end{tabular}
        };
        \end{tikzpicture}
    \captionsetup{width=0.9\linewidth}
    \caption{Running time of market clearance}
    \label{Fig:Computation_time}
 \end{figure}
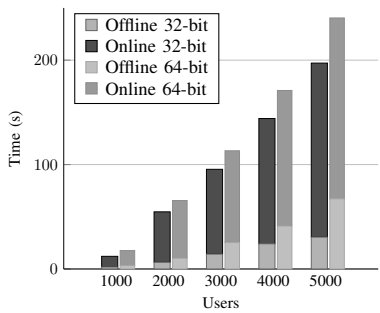
 
\textit{Bid submission.} For each trading period, each HEMS controller performs $\mathsf{Share}$ for every bid element ($4 \times \mathsf{Share}$) and a scalar multiplication to generate the Schnorr commitment, taking around $0.82$ ms.

\textit{Market clearance preprocessing.}
Each HEMS controller performs one modular multiplication to generate the Schnorr proof (negligible). Each LEMO server performs $N \times |\mathcal{P}'| \times (\mathsf{Verify} + \mathsf{Dec})$, where $\mathsf{Verify}$ and $\mathsf{Dec}$ cost $0.024~\text{ms}$ and $1.33~\mu\text{s}$, respectively, as well as $(2 \times N \times |\mathcal{P}'| ) + 1$ scalar multiplications for Schnorr batch verification, where a single scalar multiplication costs $0.52~\text{ms}$. We summarise in Table~\ref{tab:preprocessing_time} the runtimes for the preprocessing operations for different values of $N$, with $|\mathcal{P}'| = 3$.

\textit{Market clearance energy allocation.}
Most of the underlying MPC protocols used in this work divide the computation into data-independent (offline) and data-dependent (online) phases. In the former, correlated randomness is pre-generated (e.g., edaBits for comparison) to be used by the latter for the actual computation on private data. We present in Table~\ref{tab:Allocation_time} the results of the market clearance energy allocation (for 64-bit inputs) in terms of runtimes for each phase, including computation and communication costs, the number of rounds for each phase, and the combined CPU time of both phases. Figure~\ref{Fig:Computation_time} visualises the runtimes for 32-bit and 64-bit inputs.

We first note that the asymptotic behaviour of the results seems to align with the theoretical complexity analysis of Algorithms~\ref{alg:Peer_Mappings2} and~\ref{alg:MarketClearance2}, where $|\mathcal{P}'|$ is constant and $|\mathbf{B}^s| = |\mathbf{B}^b| = N/2$. We also note that CPU time represents less than $7\%$ of the overall time, which means that communication cost accounts for the majority of the execution time. Importantly, the results show that for $5{,}000$ users, the online phase of TP calculation and energy allocation can be executed in around $3$~min, and around $4$~min when including the offline phase. Adding the preprocessing time ($\approx 17$~s), the entire market clearance stage can still be completed in under $5$~min.~This indicates that PNF-LEM is practical for a typical LEM instance of a few thousand users and a trading period of $30$~min.

Despite this, it is important to recall that the number of selected and matched peers was fixed in our experiments. Based on our experiments, we observed that the majority of the energy allocation runtime is dominated by Algorithm~\ref{alg:MarketClearance2}, which grows quadratically with $|\mathcal{P}'|$. Although this indicates that $|\mathcal{P}'|$ can significantly influence the overall performance, the high runtime of Algorithm~\ref{alg:MarketClearance2} primarily results from the expensive secure comparison operations, which can be improved in practice. For instance, more efficient comparison protocols have been proposed, such as~\cite{Makri}, which achieves up to a 2.3× speedup over the protocol used in our implementation~\cite{Escudero}. Alternatively, similar to Algorithm~\ref{alg:Peer_Mappings2}, bids’ volumes can be masked as in~\cite{Mehrdad}, enabling comparisons to be be executed in the clear, which would significantly reduce the overall runtime.

\section{Conclusion}\label{sec:conclusion}
We introduced a privacy-preserving protocol for LEMs that accounts for network fees, protects participants’ privacy using MPC, and incorporates an authentication mechanism based on Schnorr identification. We optimised the protocol’s efficiency, implemented it, and analysed its performance, demonstrating its feasibility for a typical LEM scale. 
Future work will focus on addressing linkability concerns arising from the reuse of user-bound transaction fee tuple shares and extending the protocol to support security against malicious LEMO servers.

\bibliographystyle{IEEEtran}
\bibliography{Refs}

\newpage

\vfill

\end{document}